# Asymptotic Theory for Directed Transport of Suspended Ferromagnetic Nanoparticles

S.I. Denisov[1][*], T.V. Lyutyy[2][†], M.M. Moskalenko[2], A.T. Liutyi[2], and Yu.S. Bystryk[1][‡]

[1] Institute of Applied Physics, National Academy of Sciences of Ukraine,
Petropavlivska Street 58, 40000 Sumy, Ukraine

[2] Sumy State University, Kharkivska Street, 116, 40007 Sumy, Ukraine

Using the rigid dipole model, we study the translational and rotational motions of single-domain ferromagnetic nanoparticles in a dilute suspension induced by the harmonically oscillating gradient magnetic field in the presence of a time-independent uniform magnetic field. Our approach is based on a set of the first-order differential equations that describe the time dependencies of the particle coordinate and magnetization angle. We find the asymptotic solutions of this set of equations at small and large times and, by applying the matched asymptotic expansions for discrete times, derive analytical expressions for the average particle coordinate and velocity.

## 1. INTRODUCTION

Ferromagnetic nanoparticles in a single-domain state are widely used in many biomedical applications [1], such, e.g, as magnetic hyperthermia [2,3], magnetic cell separation [4,5], and drug delivery [6,7]. Usually, these applications utilize the magnetic and mechanical dynamics of nanoparticles. If the anisotropy magnetic field exceeds all external magnetic fields, then the magnetization vector can approximately be considered as 'frozen' into the particle body. In this rigid dipole model [8] the magnetization dynamics is defined by the translational and rotational dynamics of nanoparticles and thus can be excluded from consideration. This fact essentially simplifies the theoretical analysis of the nanoparticle dynamics in a viscous liquid and often permits to obtain analytical results.

Using the rigid dipole approximation, we have performed a comprehensive analysis of the coupled translational and rotational dynamics of nanoparticles subjected to a time-independent gradient magnetic field and a constant uniform magnetic field [9]. It has been shown that, due to coupling, there exist four regimes of directed transport of nanoparticles, which depend on their initial positions. If, as it often happens, the gradient magnetic field depends on time, then the coupled translational and rotational dynamics of nanoparticles becomes very complex and can lead to unexpected results. Specifically, if the gradient magnetic field changes with time harmonically, then, in accordance with our expectations, all nanoparticles perform periodic oscillations [10]. However, the application of a constant uniform magnetic field, whose direction is perpendicular to the gradient one, leads to the directed transport (drift motion) of nanoparticles [11]. This is a rather unexpected result because a uniform magnetic field does not produce any driving force. Our numerical analysis have shown that this is a dynamic effect, under which the translational oscillations become aperiodic and the nanoparticles are shifted during each period of the gradient field along (or opposite) its direction. This mechanism of directed transport has recently been analyzed both numerically and analytically [12].

In this paper, using the rigid dipole approximation, we develop an asymptotic theory of directed transport of ferromagnetic nanoparticles, which permits to derive analytic formulas for the average particle position and velocity as functions of discrete time.

## 2. BASIC ASSUMPTIONS AND EQUATIONS

We consider spherical ferromagnetic nanoparticles of radius $a$ in dilute suspension, when the dipole-dipole interaction between nanoparticles can be neglected. It is assumed that on nanoparticles act the harmonically oscillating gradient magnetic field

$$\mathbf{H}_g = gx \sin(\Omega t + \phi)\, \mathbf{e}_x \qquad (1)$$

and the time-independent uniform magnetic field

$$\mathbf{H} = H_\parallel \mathbf{e}_x + H_\perp \mathbf{e}_y. \qquad (2)$$

Here, $g(\geq 0)$ is the gradient of the magnetic field (1), $\Omega$ is its frequency, $\phi \in [0,\pi]$ is its initial phase, $H_\parallel$ and $H_\perp(\geq 0)$ are the parallel and perpendicular components of the uniform magnetic field (2), and $\mathbf{e}_x, \mathbf{e}_y, \mathbf{e}_z$ are the unit vectors along the correspondent axes of the Cartesian system of coordinates.

We also assume that nanoparticles are single-domain and the anisotropy magnetic field in them is so strong that the magnetization vector $\mathbf{M} = \mathbf{M}(t)$ ($|\mathbf{M}| = M = $ const) can be considered as 'frozen' into their bodies. This so-called rigid dipole model is widely used, e.g., for studying the role of the magnetic dipolar interaction and thermal fluctuations in the nanoparticle dynamics [13-17]. If the size of nanoparticles is rather large ($a \gtrsim 50$ nm), then thermal fluctuations of the magnetization direction are small and the dynamics of $\mathbf{M}$ can be considered in the deterministic approximation. In this case, taking the initial magnetization $\mathbf{M}_0 = \mathbf{M}(0)$ in the $xy$ plane, the magnetization vector $\mathbf{M}$ can be represented in the form

$$\mathbf{M} = M(\cos\varphi\, \mathbf{e}_x + \sin\varphi\, \mathbf{e}_y), \qquad (3)$$

where $\varphi = \varphi(t)$ is the magnetization angle (angle between the vectors $\mathbf{e}_x$ and $\mathbf{M}$). Introducing the particle angular velocity $\boldsymbol{\omega} = \boldsymbol{\omega}(t)$, we make sure that in the rigid dipole approximation the kinematic equation

$$\frac{d}{dt}\mathbf{M} = \boldsymbol{\omega} \times \mathbf{M} \qquad (4)$$

---

[*] denisov@sumdu.edu.ua
[†] lyutyytv@gmail.com
[‡] yurabystrik@gmail.com

(the sign $\times$ denotes the vector product) must hold. From (3) and (4) it follows that $\boldsymbol{\omega} = \omega_z \mathbf{e}_z$ (the nanoparticle rotates about the axis $z$) and $\omega_z = d\varphi/dt$.

Next, neglecting the inertial effects and assuming that the Reynolds rotational and translational numbers are small [18], from the torque and force balance equations we, respectively, obtain

$$\frac{d\varphi}{dt} = \frac{MV}{6\eta}[H_\perp \cos\varphi - H_\parallel \sin\varphi - gR_x \sin\varphi \sin(\Omega t + \phi)]. \quad (5)$$

($\eta$ is the dynamic viscosity of liquid, $V = (4/3)\pi a^3$ is the nanoparticle volume) and

$$\frac{dR_x}{dt} = \frac{2Mga^2}{9\eta}\cos\varphi \sin(\Omega t + \phi) \qquad (6)$$

($R_x = R_x(t)$ is the $x$-coordinate of the particle center). Finally, introducing the dimensionless time $\tau = \Omega t$, particle coordinate $r_x = r_x(\tau) = R_x/a$ and frequencies

$$\nu_g = \frac{Mga}{6\eta\Omega}, \quad \nu_\perp = \frac{MH_\perp}{6\eta\Omega}, \quad \nu_\parallel = \frac{MH_\parallel}{6\eta\Omega}, \qquad (7)$$

Eqs. (5) and (6) can be reduced to a set of the dimensionless equations for $\varphi$ and $r_x$ [11]

$$\dot\varphi = \nu_\perp \cos\varphi - \nu_\parallel \sin\varphi - \nu_g r_x \sin\varphi \sin(\tau + \phi), \quad (8)$$

$$\dot r_x = (4/3)\nu_g \cos\varphi \sin(\tau + \phi), \qquad (9)$$

where the overdot denotes the derivative with respect to the dimensionless time $\tau$. Despite the seeming simplicity of these equations, they are difficult to solve not only analytically, but also numerically. The reason is that Eqs. (8) and (9) are stiff [19], i.e., certain numerical methods for their solution become numerically unstable with time (because $|r_x(\tau)|$ infinitely grows and $\varphi(\tau)$ tends to the step function as $\tau \to \infty$, see below).

Equations (8) and (9), supplemented be the initial conditions

$$\varphi(0) = \varphi_0 \in [0,\pi], \qquad r_{x0} = r_x(0) \in (-\infty,\infty), \quad (10)$$

describe the coupled rotational and translational motions of nanoparticles in a viscous liquid subjected to the harmonically oscillating gradient magnetic field in the presence of a constant uniform magnetic field. In Refs. [9-12] we studied numerically and analytically the dynamical properties of nanoparticles on rather short time intervals. Here, we use these equations, the discrete-time approximation and matched asymptotic expansions at short and large times to derive the average particle coordinate and velocity for arbitrary times.

## 3. ASYMPTOTIC THEORY OF DIRECTED TRANSPORT

### 3.1 Nanoparticle dynamics near the origin

Let us first consider the nanoparticle dynamics near the origin (where the gradient field equals zero, i.e., $x = 0$) under the conditions

$$\nu_g \ll 1, \quad \nu_\perp \sim |\nu_\parallel| \gg 1, \quad |r_x| \ll (\nu_\perp^2 + \nu_\parallel^2)^{1/2}/\nu_g. \quad (11)$$

According to the results of Ref. [12] generalized to $\nu_\parallel \neq 0$, in this case the nanoparticle dynamics is characterized by two regimes, fast and slow. The first one occurs at $0 < \tau < \tau_{\rm tr} \ll 1$ and the second at $\tau \geq \tau_{\rm tr}$, where the transient time $\tau_{\rm tr}$ is defined as

$$\tau_{\rm tr} = \frac{\psi - \varphi_0}{(\nu_\perp^2 + \nu_\parallel^2)^{1/2}\sin(\psi - \varphi_0)} \qquad (12)$$

with

$$\psi = \arccos\frac{\nu_\parallel}{(\nu_\perp^2 + \nu_\parallel^2)^{1/2}}. \qquad (13)$$

As follows from Eqs. (8) and (9), the nanoparticle dynamics at $\tau \in (0, \tau_{\rm tr})$ is described by formulas

$$\varphi = \varphi_0 + [(\nu_\perp^2 + \nu_\parallel^2)^{1/2}\sin(\psi - \varphi_0) - \nu_g r_{x0}\sin\varphi_0 \sin\phi]\tau \qquad (14)$$

and

$$r_x = r_{x0} + (4/3)\nu_g \cos\varphi_0 \sin\phi\, \tau. \qquad (15)$$

In contrast, if $\tau \geq \tau_{\rm tr}$, then Eqs. (8) and (9) in the main approximation yield

$$\varphi = \varphi^{(0)} - r_{x0}\frac{\nu_g \nu_\perp}{\nu_\perp^2 + \nu_\parallel^2}[\sin(\tau + \phi) - \sin(\tau_{\rm tr} + \phi)] \quad (16)$$

and

$$r_x = r_x^{(0)} + r_{x0}\frac{2\nu_g^2 \nu_\perp^2}{3(\nu_\perp^2 + \nu_\parallel^2)^{3/2}}(\tau - \tau_{\rm tr}). \qquad (17)$$

Here, the constants $\varphi^{(0)}$ and $r_x^{(0)}$ can be found from (14) and (15) as $\varphi^{(0)} = \varphi(\tau_{\rm tr})$ and $r_x^{(0)} = r_x(\tau_{\rm tr})$. It should also be emphasized that the right-hand sides of Eqs. (14)–(17) are approximated (e.g., in (17) we do not show small periodic terms).

Next, we will be interested in the behavior of $r_x$ at discrete times $\tau_n = 2\pi n + \tau_{\rm tr}$, where $n = 1, 2, \dots, N$ and $N$ is the maximal number, such that the representation (17) still holds. To derive the formula for $r_x(2\pi n)$, we first introduce the nanoparticle average velocity $\bar v_n$ on the $n$-th period of the gradient magnetic field as

$$\bar v_n = \frac{1}{2\pi}[r_x(2\pi n + \tau_{\rm tr}) - r_x(2\pi n - 2\pi + \tau_{\rm tr})]. \quad (18)$$

Since $\tau_{\rm tr} \ll 1$, the dimensionless particle coordinate $r_x(2\pi n + \tau_{\rm tr})$ can be written in the form

$$r_x(2\pi n + \tau_{\rm tr}) = \begin{cases} r_x(2\pi n), & n = 1,2,\dots,N, \\ r_x(\tau_{\rm tr}) \approx r_{x0}, & n = 0, \end{cases} \qquad (19)$$

This result permits us redefine the average velocity (18) as follows:

$$\bar{v}_n = \frac{1}{2\pi}[r_x(2\pi n) - r_x(2\pi n - 2\pi)]. \quad (20)$$

Generalizing results derived in [12] to the case with $v_\parallel \neq 0$, we obtain

$$\bar{v}_n = r_x(2\pi n - 2\pi)\frac{2v_g^2 v_\perp^2}{3(v_\perp^2 + v_\parallel^2)^{3/2}}. \quad (21)$$

Using (21) together with the definition (20), for the nanoparticle coordinate $r_x(2\pi n)$ one gets

$$r_x(2\pi n) = r_x(2\pi n - 2\pi)\left(1 + \frac{4\pi v_g^2 v_\perp^2}{3(v_\perp^2 + v_\parallel^2)^{3/2}}\right) \quad (22)$$

or

$$r_x(2\pi n) = r_{x0}\left(1 + \frac{4\pi v_g^2 v_\perp^2}{3(v_\perp^2 + v_\parallel^2)^{3/2}}\right)^n. \quad (23)$$

Thus, the particle coordinate grows with $n$ in accordance with the power law.

### 3.2 Nanoparticle dynamics far from the origin

For nanoparticles far from the origin the conditions (11) should be replaced by

$$v_g \ll 1, \quad v_\perp \sim |v_\parallel| \gg 1, \quad |r_x| \gtrsim (v_\perp^2 + v_\parallel^2)^{1/2}/v_g. \quad (24)$$

According to (8) and (9), in this case $\cos\varphi$ as a function of $\tau$ tends to the periodic step function

$$\cos\varphi = \pm \begin{cases} 1 - \left(\frac{v_\perp}{v_g r_{x0} + v_\parallel}\right)^2, & 0 < \tau < \pi - \phi, \\ -1 + \left(\frac{v_\perp}{v_g r_{x0} - v_\parallel}\right)^2, & \pi - \phi < \tau < 2\pi - \phi, \\ 1 - \left(\frac{v_\perp}{v_g r_{x0} + v_\parallel}\right)^2, & 2\pi - \phi < \tau < 2\pi. \end{cases} \quad (25)$$

Here, the signs '+' and '−' in the symbol '±' correspond to $r_{x0} > 0$ and $r_{x0} < 0$, respectively. For such nanoparticles the average velocity

$$\bar{v}_\infty = \frac{2v_g}{3\pi} \lim_{N\to\infty} \int_{2\pi N - 2\pi}^{2\pi N} \cos\varphi \sin(\tau + \phi)\, d\tau \quad (26)$$

can be represented as

$$\bar{v}_\infty = \frac{2v_g}{3\pi} \int_0^{2\pi} \cos\varphi \sin(\tau + \phi)\, d\tau. \quad (27)$$

Substituting (25) into (27) and performing the integration assuming that $v_g|r_{x0}| \gg |v_\parallel|$, one obtains

$$\bar{v}_\infty = \pm\frac{8 v_g}{3\pi}\left[1 - \left(\frac{v_\perp}{v_g r_{x0}}\right)^2\right]. \quad (28)$$

This result shows that if $|r_{x0}|$ is rather large (or time $\tau$ is so large that $|r_x(\tau)| \gtrsim (v_\perp^2 + v_\parallel^2)^{1/2}/v_g$), then the nanoparticles move with a constant average velocity. Therefore, the nanoparticle coordinates $r_x(2\pi n)$ at discrete times $\tau_n = 2\pi n$ are given by

$$r_x(2\pi n) = r_x(2\pi L) + 2\pi \bar{v}_\infty(n - L), \quad (29)$$

where $n = L, L+1, \ldots$ and $|r_x(2\pi L)| \gtrsim (v_\perp^2 + v_\parallel^2)^{1/2}/v_g$.

### 3.3 Matched asymptotic analysis

To find the nanoparticle coordinates for arbitrary discrete times, we will use the asymptotic formulas (23) and (29), which describe the behavior of $r_x(2\pi n)$ for $n \leq N$ and $n \geq L$, respectively. The matched asymptotic analysis [19] allows us to do this in the simplest way. For this we assume that there is a such number $N$ of the gradient field periods that the average nanoparticle velocity $\bar{v}_N$ coincides with the limiting velocity $\bar{v}_\infty$ (i.e., $N = L$). Using (20), (23) and (28), the condition $\bar{v}_N = \bar{v}_\infty$, which we consider as the equation for determining $N$, can be rewritten in the form

$$\left(1 + \frac{4\pi v_g^2 v_\perp^2}{3(v_\perp^2 + v_\parallel^2)^{3/2}}\right)^{N-1} = \frac{4(v_\perp^2 + v_\parallel^2)^{3/2}}{\pi v_g v_\perp^2 |r_{x0}|}. \quad (30)$$

Since in the left-hand side of this equation the second term within the brackets is small compared to 1 and its right-hand side exceeds 1 [see (24)], Eq. (30) yields

$$N = \frac{3(v_\perp^2 + v_\parallel^2)^{3/2}}{4\pi v_g^2 v_\perp^2} \ln\left(\frac{4(v_\perp^2 + v_\parallel^2)^{3/2}}{\pi v_g v_\perp^2 |r_{x0}|}\right). \quad (31)$$

Finally, using (23), (29) and (31), for the nanoparticle coordinates at discrete times $\tau_n = 2\pi n$ we find

$$r_x(\tau_n) = \begin{cases} r_{x0}\left(1 + \frac{4\pi v_g^2 v_\perp^2}{3(v_\perp^2 + v_\parallel^2)^{3/2}}\right)^n, & n = 1, \ldots, N, \\ r_x(2\pi N) + 2\pi \bar{v}_\infty(n - N), & n = N+1, \ldots \end{cases} \quad (32)$$

To verify this analytical result, we solve the basic Eqs. (8) and (9) numerically. We consider the SmCo$_5$ nanoparticles [21] suspended in water at room temperature (293 K) and characterized by the following parameters: $M = 1.36 \times 10^3$ emu cm$^{-3}$, $\eta = 1.00 \times 10^{-2}$ P and $a = 3 \times 10^2$ nm $= 3 \times 10^{-5}$ cm. It is also assumed that the gradient magnetic field (1) is defined by the parameters $g = 10^2$ Oe cm$^{-1}$, $\Omega = 4 \times 10^2$ rad s$^{-1}$, $\phi = \pi/2$ rad, and the uniform magnetic field has two equal components with $H_\parallel = H_\perp = 10$ Oe. For this case, the dimensionless parameters in Eqs. (8) and (9) are given by $v_g \approx 1.67 \times 10^{-1}$ and $v_\perp = v_\parallel \approx 5.56 \times 10^2$. Choosing the initial angle to be $\varphi_0 = \pi/3$ rad, from (13) and (12) we find $\psi = \pi/4$ rad and $\tau_{tr} \approx 3.15 \times 10^{-4}$, respectively. Finally, to estimate the time-step $\Delta\tau$ and the time domain $(0, \tau_{max})$ for these parameters, we should take into account that Eqs. (8) and (9) are stiff. This means that $\Delta\tau$ must be so small (compared to the gradient field period $2\pi$) that the solutions of these equations remain numerically stable on the whole time domain. On the other side, $\Delta\tau$ must not be too small in order to make the simulation time restricted by a few hours. According to (31), in the considered case we have

$$N \approx 1.35 \times 10^4 \ln\left(\frac{1.20 \times 10^4}{|r_{x0}|}\right). \quad (33)$$

Therefore, $\tau_{max}$ must satisfy the condition $\tau_{max} \gtrsim 2\pi N$.

In Fig. 1, we show by the blue line the theoretical dependence (32) of the particle coordinate $r_x(\tau_n)$ on the discrete time $\tau_n = 2\pi n$ for the system parameters introduced above and $|r_{x0}| = 10^2$. As it follows from (33), in this cases $N \approx 6.46 \times 10^4$ ($\tau_N \approx 4.06 \times 10^5$). Next, using the definition (20) of the average transport velocity on the $n$-th period of the gradient magnetic field and formula (32) for the nanoparticle coordinate, one gets

$$\bar{v}_n = \frac{2v_g^2 v_\perp^2 r_{x0}}{3(v_\perp^2 + v_\parallel^2)^{3/2}} \left(1 + \frac{4\pi v_g^2 v_\perp^2}{3(v_\perp^2 + v_\parallel^2)^{3/2}}\right)^{n-1}, \quad (34)$$

for $n = 1, ..., N$ and $\bar{v}_n = \bar{v}_\infty$ for $n \geq N + 1$. The numerical results for $r_x(\tau_n)$ and $\bar{v}_n$, obtained by the numerical solution of Eqs. (8) and (9), are shown in Figs. 1 by the red circle symbols. As seen, the theoretical and numerical results are in a qualitative agreement with each other. Some difference between these results at large dimensionless times appears because for $\bar{v}_\infty$ was used only the main term of the asymptotic expansion.

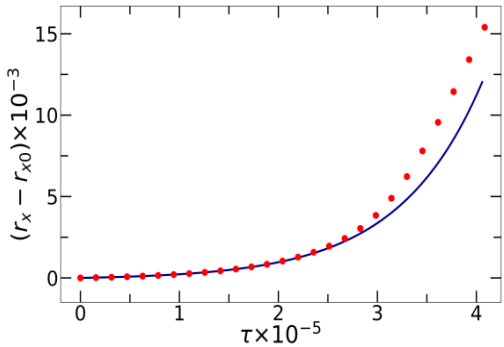

**Fig. 1** – The dependency of the particle coordinate $r_x$ on the discrete time $\tau_n = 2\pi n$ for SmCo$_5$ nanoparticles with the parameters $M = 1.36 \times 10^3$ emu cm$^{-3}$, $\eta = 1.00 \times 10^{-2}$ P, $a = 3 \times 10^2$ nm. The gradient magnetic field is defined by the parameters $g = 10^2$ Oe cm$^{-1}$, $\Omega = 4 \times 10^2$ rad s$^{-1}$, $\phi = \pi/2$ rad, and the uniform magnetic field has two equal components with $H_\parallel = H_\perp = 10$ Oe.

## 4. CONCLUSIONS

We have proposed an asymptotic theory of directed transport of single-domain ferromagnetic nanoparticles in a dilute suspension subjected to the harmonically oscillating gradient magnetic field and uniform magnetic field with two components, perpendicular and parallel to the gradient one. Using the rigid dipole approximation, we have derived a basic set of two ordinary differential equations of the first order for the time-dependent magnetization angle and dimensionless nanoparticle coordinate. By solving this set of equations for nanoparticles near and far from the origin, the asymptotic expansions of their solutions have been found at discrete times for nanoparticles near and far from the origin of the gradient magnetic field, or at small and large discrete times, respectively.

### ACKNOWLEDGEMENTS

The authors have received funding through the EURIZON project (grant agreement No. EU–3056), which is funded by the European Union under grant agreement No. 871072. The authors also acknowledge the IEEE Magnetic Society for support within the program "Magnetism for Ukraine 2022", and the Ministry of Education and Science of Ukraine for support in part by the budget program for KPKVK 2201390 (2022).

# Асимптотична теорія спрямованого транспорту зважених феромагнітних наночастинок

С.І. Денисов[1], Т.В. Лютий[2], М.М. Москаленко[2], А.Т. Лютий[2], Ю.С. Бистрик[1]

[1] *Інститут прикладної фізики НАН України,*
*вул. Петропавлівська, 58, 40000 м. Суми, Україна*

[2] *Сумський державний університет, вул. Харківська, 116, 40007 м. Суми, Україна*

В роботі представлені результати асимптотичної теорії спрямованого транспорту однодоменних феромагнітних наночастинок, що індукується градієнтним магнітним полем, яке змінюється у часі за гармонічним законом, та постійним однорідним магнітним полем, яке має перпендикулярну та паралельну до напрямку градієнтного поля компоненти. Розв'язано систему базових рівнянь, що описують обертальний та поступальний рухи наночастинок, що знаходяться в околі початку координат і знайдено їх асимптотичну поведінку. Наближений розв'язок отримано і в другому граничному випадку, коли частинки знаходяться далеко від початку координат. Використовуючи узгоджений асимптотичний аналіз, визначено часові залежності координати частинки і її середньої швидкості, які якісно узгоджуються з чисельними результатами.

**Ключові слова:** Феромагнітні наночастинки, Розведені суспензії, Градієнтне та однорідне магнітні поля, Трансляційна та обертальна динаміка, Спрямований транспорт, Узгоджений асимптотичний аналіз.